\documentclass{aastex63}

\shorttitle{GW190426: the NSBH or BBH merger event?}
\shortauthors{Li et al.}

\begin{document}

\title{GW190426\_152155: a merger of neutron star-black hole or low mass binary black holes?}


\author[0000-0001-5087-9613]{Yin-Jie Li}
\affiliation{Key Laboratory of Dark Matter and Space Astronomy, Purple Mountain Observatory, Chinese Academy of Sciences, Nanjing 210023, Peoples Republic of China}
\affiliation{School of Astronomy and Space Science, University of Science and Technology of China, Hefei, Anhui 230026, Peoples Republic of China}

\author[0000-0001-9034-0866]{Ming-Zhe Han}
\affiliation{Key Laboratory of Dark Matter and Space Astronomy, Purple Mountain Observatory, Chinese Academy of Sciences, Nanjing 210023, Peoples Republic of China}
\affiliation{School of Astronomy and Space Science, University of Science and Technology of China, Hefei, Anhui 230026, Peoples Republic of China}

\author[0000-0001-9120-7733]{Shao-Peng Tang}
\affiliation{Key Laboratory of Dark Matter and Space Astronomy, Purple Mountain Observatory, Chinese Academy of Sciences, Nanjing 210023, Peoples Republic of China}
\affiliation{School of Astronomy and Space Science, University of Science and Technology of China, Hefei, Anhui 230026, Peoples Republic of China}

\author[0000-0001-9626-9319]{Yuan-Zhu Wang}
\affiliation{Key Laboratory of Dark Matter and Space Astronomy, Purple Mountain Observatory, Chinese Academy of Sciences, Nanjing 210023, Peoples Republic of China}

\author[0000-0002-7869-0174]{Yi-Ming Hu}
\affiliation{TianQin Research Center for Gravitational Physics \& School of Physics and Astronomy, Sun Yat-sen University, 2 Daxue Road, Zhuhai 519082, Peoples Republic of China}

\author[0000-0003-4891-3186]{Qiang Yuan}
\affiliation{Key Laboratory of Dark Matter and Space Astronomy, Purple Mountain Observatory, Chinese Academy of Sciences, Nanjing 210023, Peoples Republic of China}
\affiliation{School of Astronomy and Space Science, University of Science and Technology of China, Hefei, Anhui 230026, Peoples Republic of China}

\author[0000-0002-8966-6911]{Yi-Zhong Fan}
\affiliation{Key Laboratory of Dark Matter and Space Astronomy, Purple Mountain Observatory, Chinese Academy of Sciences, Nanjing 210023, Peoples Republic of China}
\affiliation{School of Astronomy and Space Science, University of Science and Technology of China, Hefei, Anhui 230026, Peoples Republic of China}
\email{The corresponding author: yzfan@pmo.ac.cn (Y.Z.F)}

\author[0000-0002-9758-5476]{Da-Ming Wei}
\affiliation{Key Laboratory of Dark Matter and Space Astronomy, Purple Mountain Observatory, Chinese Academy of Sciences, Nanjing 210023, Peoples Republic of China}
\affiliation{School of Astronomy and Space Science, University of Science and Technology of China, Hefei, Anhui 230026, Peoples Republic of China}


\begin{abstract}
GW190426\_152155 was recently reported as one of the 39 candidate gravitational wave (GW) events in \citet{2020arXiv201014527A}, which has an unusual source-frame chirp mass $\sim 2.4M_{\odot}$ and may be the first GW signal from a neutron star-black hole (NSBH) merger. Assuming an astrophysical origin, we reanalyze GW190426\_152155 using several waveforms with different characteristics, and consider two different priors for the mass ratio of the binary (Uniform and LogUniform). We find that the results are influenced by the priors of mass ratio, and this candidate could also be from the merger of two low mass black holes (BH). 
In the case for a binary black hole (BBH) merger, the effective spin is likely negative and the effective precession spin is non-negligible.
As for the NSBH merger, supposing the mass of the light object follow the distribution of current neutron stars (NSs) with a reasonably measured/constrained mass, the spin of the low mass BH is so small   
that is hard to generate bright electromagnetic emission. Finally, we estimate a merger rate of GW190426\_152155-like systems to be $59^{+137}_{-51}~{\rm Gpc}^{-3}~{\rm yr}^{-1}$.

\end{abstract}


\section{Introduction} \label{sec:intro}

    The first three observing runs (O1, O2, O3) of the Advanced LIGO and Virgo (joined in O2) have been accomplished. The raw data of the entire O1, O2,  and 39 candidate GW events during O3a are now publicly available \citep[\url{https://www.gw-openscience.org/data};][]{2020arXiv201014527A}. Since the first detection of GW signal from a BBH coalescence in 2015 \citep{2016PhRvL.116f1102A}, $\sim$ 50 BBH events have been reported (\url{https://www.gw-openscience.org/eventapi/html/GWTC}) , which allows us to study the population properties of BBH coalescences and enhances our understandings of the stellar formation and evolution \citep{2019ApJ...882L..24A,2020arXiv201014533T}. Moreover, the first GW signal from a binary neutron star (BNS) coalescence was detected on 17th August 2017 \citep{2017PhRvL.119p1101A,2019PhRvX...9a1001A}, and its associated electromagnetic (EM) emission was also observed \citep{2017ApJ...848L..12A,2017ApJ...848L..13A}. The data of the BNS merger event GW170817 have been widely adopted to constrain the equation of state (EOS) of the ultra-dense matter as well as the bulk properties of the NSs \citep[e.g.,][]{2017PhRvL.119p1101A,2018PhRvL.121p1101A, 2018PhRvL.120q2703A, 2018PhRvL.121i1102D, 2018PhRvL.120z1103M, 2019ApJ...887L..22R, 2020ApJ...892...55J, 2020PhRvD.101l3007L}. Besides, the detections of GW from compact binary coalescences also play important roles  in testing the accuracy of General relativity (GR) \citep[e.g.,][]{2016PhRvL.116v1101A,2019PhRvL.123a1102A,2019PhRvD.100j4036A,2020arXiv201014529T}, and measuring the Hubble Constant \citep[e.g.,][]{2017Natur.551...85A,2019ApJ...876L...7S}.
    
    The NSBH binaries have been widely expected in the stellar evolution theories \citep[e.g.,][]{2006NatPh...2..116G, 2008ApJ...676.1162S, 2010ApJ...720..953L, 2013LRR....16....4B}, and they are one of the important targets of the advanced LIGO/Vigro/KARGRA detectors \citep[see][and its refences]{2020LRR....23....3A}. Indeed, the unexpectedly heavy BNS event GW190425 \citep{2020ApJ...892L...3A} is also consistent with being a NSBH merger \citep[see][]{2020ApJ...891L...5H}. Additionally, GW190814 involving a $22.2-24.3 M_{\odot}$ BH and a compact object with a mass of $2.50-2.67 M_{\odot}$ cannot be ruled out as a NSBH merger event either \citep{2020ApJ...896L..44A}.
     In addition to radiating strong GW emission, the NSBH mergers can also produce electromagnetic transients such as short/long-short gamma-ray bursts (GRBs) and kilonovae/macronovae if the merging NSs have been effectively tidally disrupted by their BH partners \citep{1992ApJ...395L..83N,1998ApJ...507L..59L,2004RvMP...76.1143P,2019LRR....23....1M}. Actually, the KN of long-short GRB060614 can be interpreted as the consequence of a NSBH merger \citep{2015NatCo...6.7323Y,2015ApJ...811L..22J}. Thus, the NSBH binary systems play an important role in studying physics, astronomy, and astrophysics.
    
    The GW190426\_152155 is one of the 39 candidate GW events reported in O3a catalog \citep{2020arXiv201014527A} with a network signal-to-noise ratio (SNR) of 10.1 and a false alarm rate (FAR) of $1.4 \rm {yr}^{-1}$, involving a component object with a mass $\textless 3 M_{\odot}$, while its astrophysical probability has not been estimated since it strongly depends on the prior assumptions of the merger rate of such binaries.  In this work we assume it as a real signal of astrophysical origin and investigate the possibility that it could have originated from either a NSBH or a BBH merger in Sec.\ref{NSBH:BBH}, and further study the properties of candidate as a NSBH or a BBH in Sec.\ref{NSBH}/Sec.\ref{BBH}. Finally we estimate the merger rate of such systems in Sec.\ref{merger rate}. 
    
\section{Data Analysis} \label{sec:PE}

As reported in \citet{2020arXiv201014527A}, GW190426\_152155 may be the first detected NSBH event. However, the BBH merger case for GW190426\_152155 cannot be ruled out, due to the uninformative measurement of the tidal deformability, which would definitively indicate whether it had a NSBH or BBH origin. In this work, we only consider the component masses to determine the origin of this system, leaving out the possibility it may contain a BH lighter than the heaviest stable NS \citep{2016PhRvD..94h3504C,2019arXiv190608217C}.

To obtain the source parameters of GW190426\_152155 we use Bayesian Inference method, applying the open source software BILBY \citep{2019ApJS..241...27A}.
We use 128s of the strain data spanning GPS time (tc\footnote{tc is the geocentric GPS time of the merger.}-120s, tc+8s) which are available from the Gravitational Wave Open Science Center (Catalog) (see \url{https://www.gw-openscience.org/eventapi/html/GWTC-2/GW190426_152155/v1/}). As for the noise power spectral density (PSD), we use the pre-estimated PSD files from LIGO Document Control Center (see \url{https://dcc.ligo.org/LIGO-P2000223/public}). To evaluate the likelihood, we use the
 data with low and high frequency cutoff at 18 and 2048 Hz. In order to investigate the two scenarios (i.e. BBH and NSBH) for GW190426\_152155, we use two groups of GW waveforms to fit the data, one group that do not consider the tidal effect, including \text{IMRPhenomXAS} \citep{2020PhRvD.102f4001P}, \text{IMRPhenomXP}  \citep{2014PhRvL.113o1101H,2015PhRvD..91b4043S,2016PhRvD..93d4006H,2016PhRvD..93d4007K,2017PhRvD..95j4004C,2020arXiv200110897G}, and \text{IMRPhenomXPHM} \citep{2018PhRvL.120p1102L,2020arXiv200110897G,2017PhRvD..95j4004C,2019PhRvD.100b4059K}. Another group including \text{IMRPhenomNSBH} \citep{2020PhRvD.101l4059T} and \text{SEOBNRv4\_ROM\_NRTidalv2\_NSBH} \citep{2020PhRvD.102d3023M} consider the tidal deformability for the secondary object, while do not take into account the procession effect and higher mode imprints. 
We do not consider the calibration errors of the detectors since it will influence the sky localization but has little effect on mass measurements \citep{2016PhRvL.116f1102A}. 

As is well known that the priors will also influence the inferred GW parameters, especially for the case that the data are not well informative. We mainly consider two priors for the compact binary masses, one is that the mass ratio ($q = M_2/M_1$) is LogUniform from 0.05 to 1 and the detector-frame chirp mass ($\mathcal{M}^{\rm det}$) is uniform in (2.55,2.65)$M_{\odot}$, following LIGO's configuration \citep{2020arXiv201014527A}; the other is that q uniformly distributes in (0.05,1), and $\mathcal{M}^{det}$ is uniform in (2.55,2.65)$M_{\odot}$. As for the other parameters, the prior is set similar to that in \citet{2020arXiv201014527A}, which is uniform in spin magnitudes, isotropic in spin orientations, sky locations, and binary orientation; the luminosity distance is set by assuming that the source is uniform in co-moving volume bounded in $(50,1000)$ Mpc, with a flat $\Lambda$CDM cosmology, where the matter density $\Omega_m=0.307$, the dark energy density $\Omega_{\rm \Lambda}=0.6925$, and the Hubble constant $H_0=67.74~{\rm km~s^{-1}~Mpc^{-1}}$ \citep{2016A&A...594A..13P}. 
Finally, we conduct all the inferences with 2000 live points by the sampler Pymultinest \citep{2016ascl.soft06005B}. 

\begin{figure*}
\centering
\includegraphics[scale=0.4]{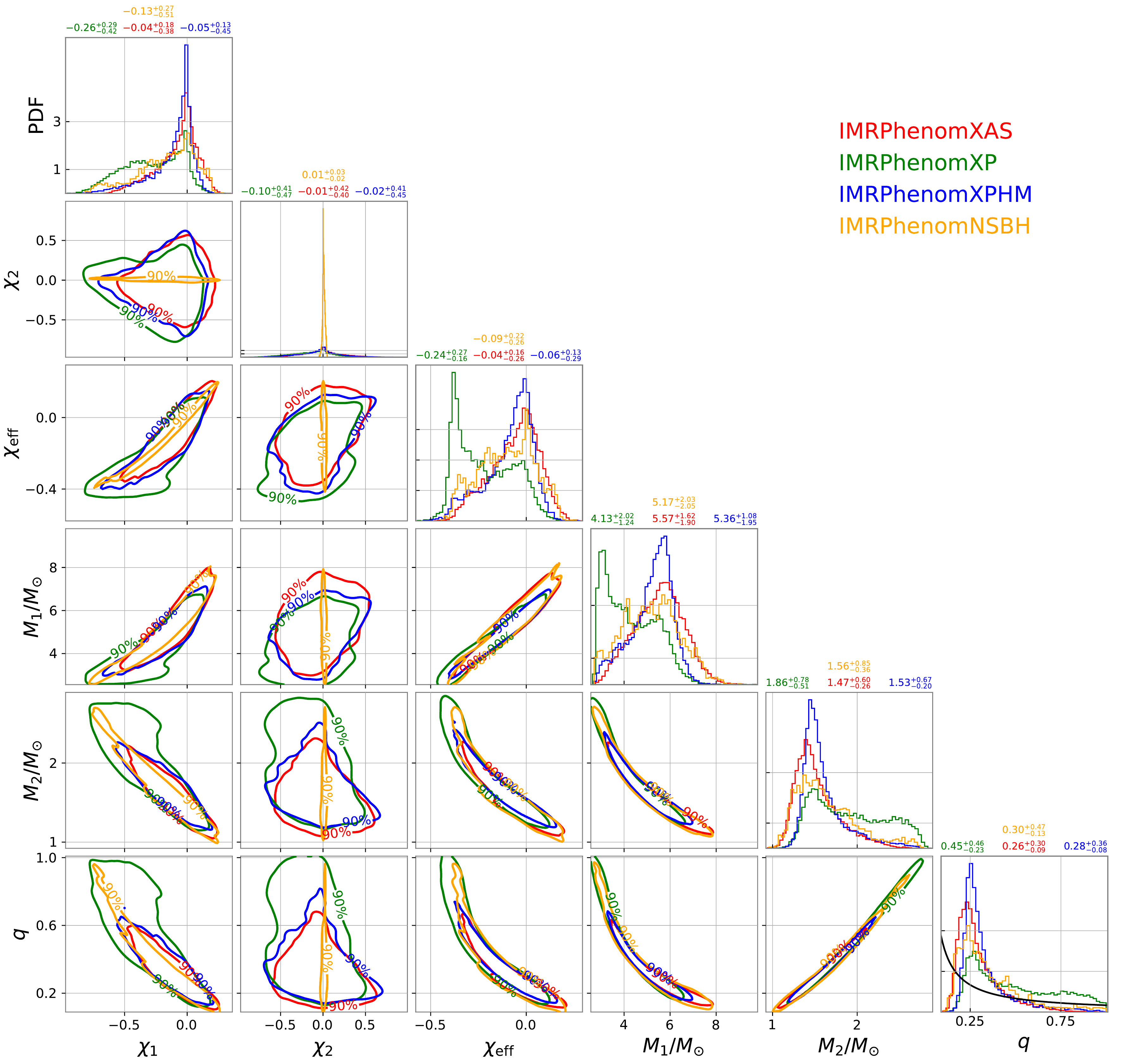}
\caption{Posterior distributions of the physical parameters of GW190426\_152155 obtained by the \text{IMRPhenomNSBH}, \text{IMRPhenomXAS}, \text{IMRPhenomXP}, and \text{IMRPhenomXPHM} waveforms with the LogUniform mass ratio prior, including the mass ratio $q$, the source frame masses of two compact objects ($M_1,~M_2$), the  aligned spins of two compact objects ($\chi_1,~\chi_2$), and the effective spin ($\chi_{\rm eff}$). The uncertainty represent the corresponding 90\% confidence interval. The black line in the sub-figure of mass ratio's marginalized distribution denotes its prior.}
\label{logq}
\end{figure*}

\subsection{The general case}\label{NSBH:BBH}

\begin{figure*}
\centering
\includegraphics[scale=0.4]{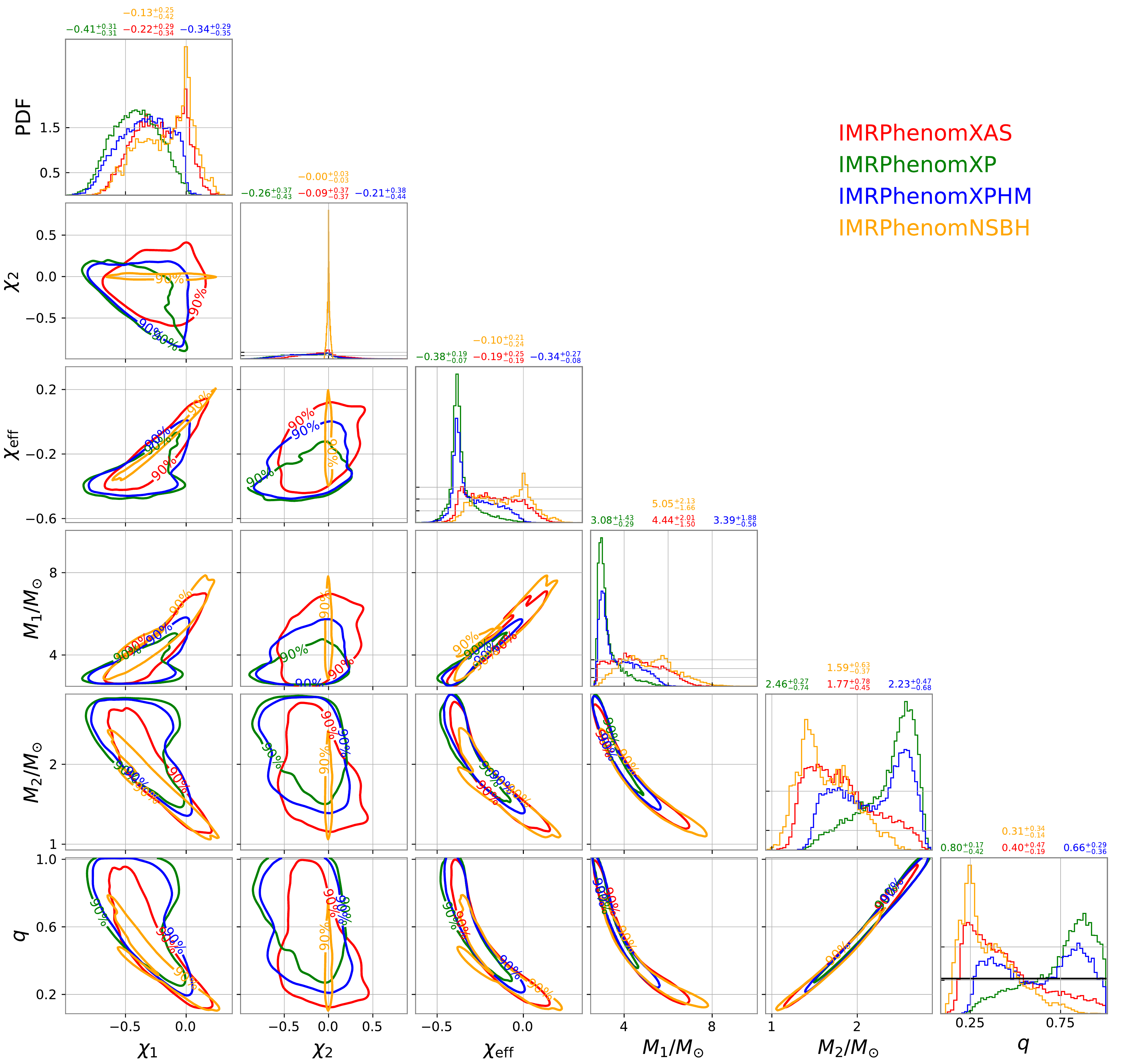}
\caption{
The same as Fig.\ref{logq} but for the flat mass ratio prior}
\label{flatq}
\end{figure*}

Three BBH waveforms are employed to extract the signal from the strain data of GW190426\_152155. \text{IMRPhenomXAS} is a frequency domain, non-precessing phenomenological inspiral-merger-ringdown (IMR) waveform model \citep{2020PhRvD.102f4001P}, while \text{IMRPhenomXP} adds the precession effect to the former \citep{2014PhRvL.113o1101H,2015PhRvD..91b4043S,2016PhRvD..93d4006H,2016PhRvD..93d4007K,2017PhRvD..95j4004C,2020arXiv200110897G}, and \text{IMRPhenomXPHM} includes both precession effect and higher mode imprints \citep{2018PhRvL.120p1102L,2020arXiv200110897G,2017PhRvD..95j4004C,2019PhRvD.100b4059K}. The spin magnitudes of both components are set uniform in (0,0.99) , i.e. the high spin (HS) case. The source parameters posteriors for the two priors mentioned above are shown in Fig.\ref{logq} and Fig.\ref{flatq}, including the mass ratio $q$, source frame component masses ($M_1$, $M_2$), aligned spins of two compact objects ($\chi_1,~\chi_2$), and the effective spin ($\chi_{\rm eff} = (M_1\chi_1+M_2\chi_2)/(M_1+M_2)$).

From the two figures, we find that the mass ratio of the system is not constrained well by either waveforms, and the priors influence the posterior distributions effectively. Consequently, the parameter space of the spins of two component objects remains to be uncertain, due to the mass ratio-spin degeneracy \citep{1994PhRvD..49.2658C,2013PhRvD..88d2002O,2013PhRvD..87b4035B}. From Fig.\ref{logq}, we find that all waveforms favor a low mass ratio mode in the posteriors of GW190426\_152155, i.e. the secondary object may have a mass below the maximum stable NS.  While for the \text{IMRPhenomXP} waveform, the posterior in the region of high mass ratio is also non negligible, and this leads to bimodal distributions of $M_1$ and $\chi_{\rm eff}$, which are different from the posterior distributions obtained by other waveforms. As for the results of the flat mass ratio prior in Fig.\ref{flatq}, \text{IMRPhenomXP} and \text{IMRPhenomXPHM} favor a high mass ratio mode, (see their posterior distributions,) effective spin ($\chi_{\rm eff}$) is significantly negative, and $M_1$ has a peak $\sim 3M_{\odot}$, which indicates that GW190426\_152155 may be a symmetric BBH system. Meanwhile, the \text{IMRPhenomXAS} still more favors the low mass ratio mode than the high mass ratio mode.

As for the NSBH model, the \text{IMRPhenomNSBH} waveform is employed, which takes into account the tidal deformability of the secondary object but ignores the precession effect and higher mode imprints, another NSBH waveform \text{SEOBNRv4\_ROM\_NRTidalv2\_NSBH} \citep{2020PhRvD.102d3023M} is used for cross check, and we find the results obtained by two waveforms are rather similar. We take the low spin case for the NS , i.e. spin magnitude of the NS is set uniform in (0,0.05), while the tidal deformability of NS ($\Lambda_2$) is uniform in (0,5000) and the detector frame secondary component mass is constrained $\textless 3M_{\odot}$ (waveform required). As shown in Fig.\ref{logq} and Fig.\ref{flatq}, the posterior distributions are similar to that obtained by \text{IMRPhenomXAS} for both mass ratio priors, this is because that \text{IMRPhenomNSBH} and \text{IMRPhenomXAS} are both non-precessing waveforms.
The tidal deformability of the NS is not constrained i.e. the results show that the posterior of $\Lambda_2$ are similar to its prior, which was also mentioned in \citet{2020arXiv201014527A}. Note that, the distribution of $\chi_2$ obtained from NSBH waveform is rather narrow, and that is caused by the constraint of the spin magnitude of the secondary object $a_1 \textless 0.05$ in prior.

\begin{figure*}
\gridline{\fig{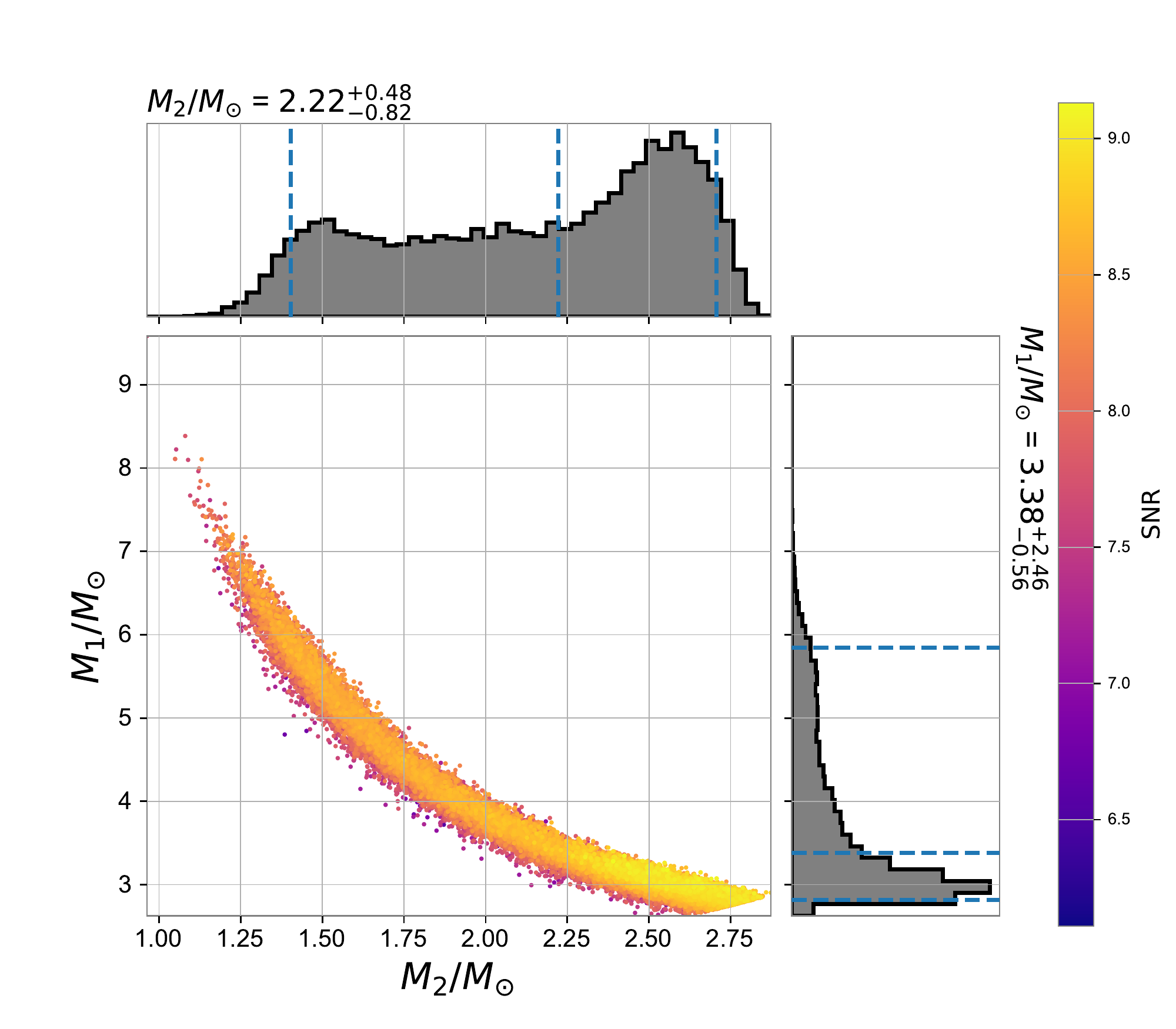}{0.4\textwidth}{(IMRPhenomXP)}
          \fig{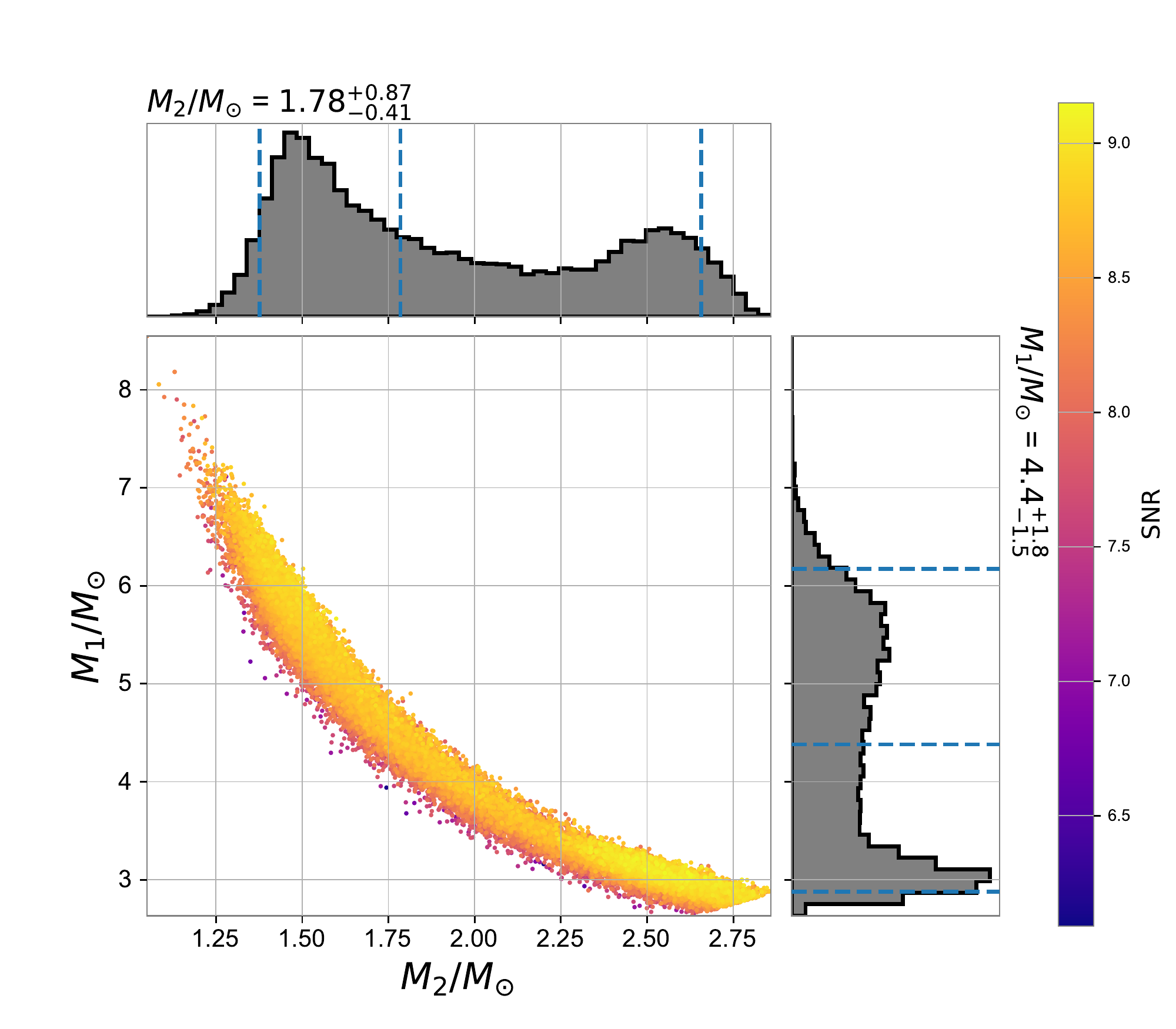}{0.4\textwidth}{(IMRPhenomXPHM)}
          }
\gridline{\fig{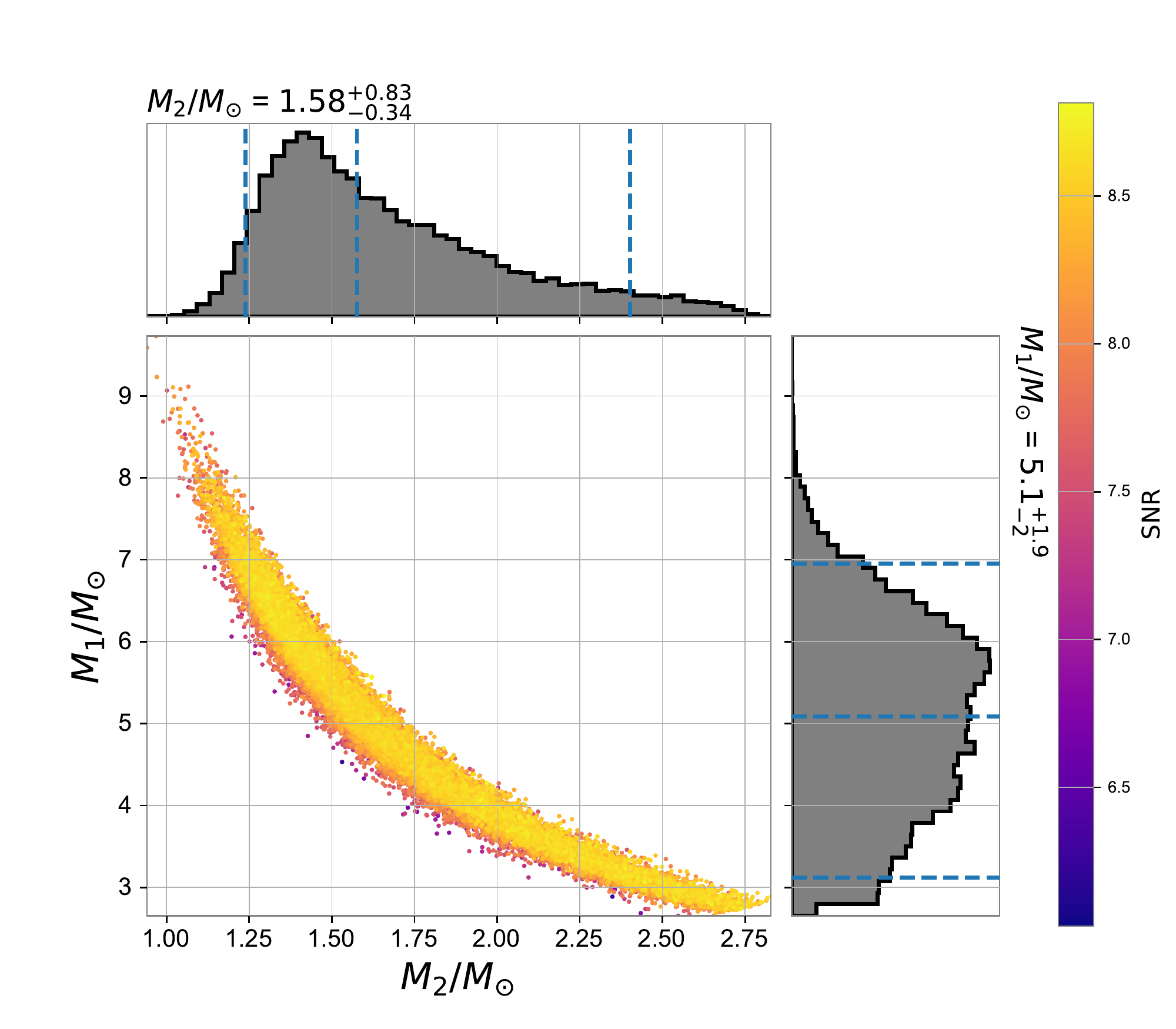}{0.4\textwidth}{(IMRPhenomXAS)}
         \fig{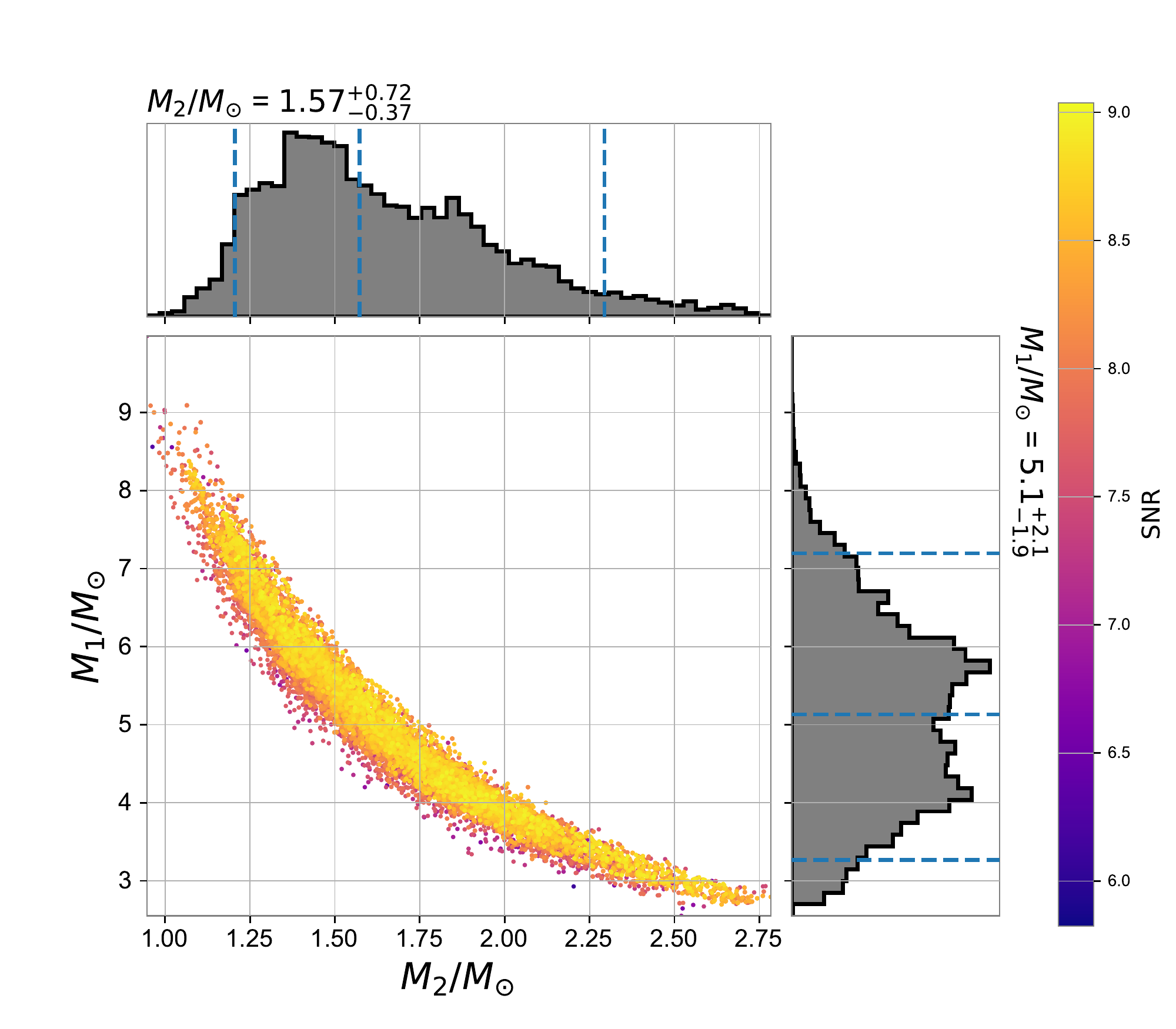}{0.4\textwidth}{(IMRPhenomNSBH)}
          }
\caption{The posteriors from the inferences of two priors cases are combined together and the samples are colored by SNR ($\sqrt{2\ln\mathcal{L}}$), where $\mathcal{L}$ is the likelihood ratio for a given point. 
}
\label{snr}
\end{figure*}

We also plot the posteriors of the component masses that colored by SNR ($\sqrt{2\ln\mathcal{L}}$), where $\mathcal{L}$ is the likelihood ratio for a given point. In order to investigate a broader parameter space and compare the two modes (i.e. the low q mode and the high q mode) directly, we combine the posteriors from both priors together. As shown in Fig.\ref{snr}, in all four results that obtained by the different waveforms, the SNRs are comparable in the whole $M_2$'s range from $\sim 1.2M_{\odot}$ to $\sim2.75M_{\odot}$. 
Therefore, we cannot rule out either the NSBH or the BBH merger origin for GW190426\_152155.

\subsection{The case of BBH merger}\label{BBH}

\begin{figure*}
\centering
\includegraphics[scale=0.3]{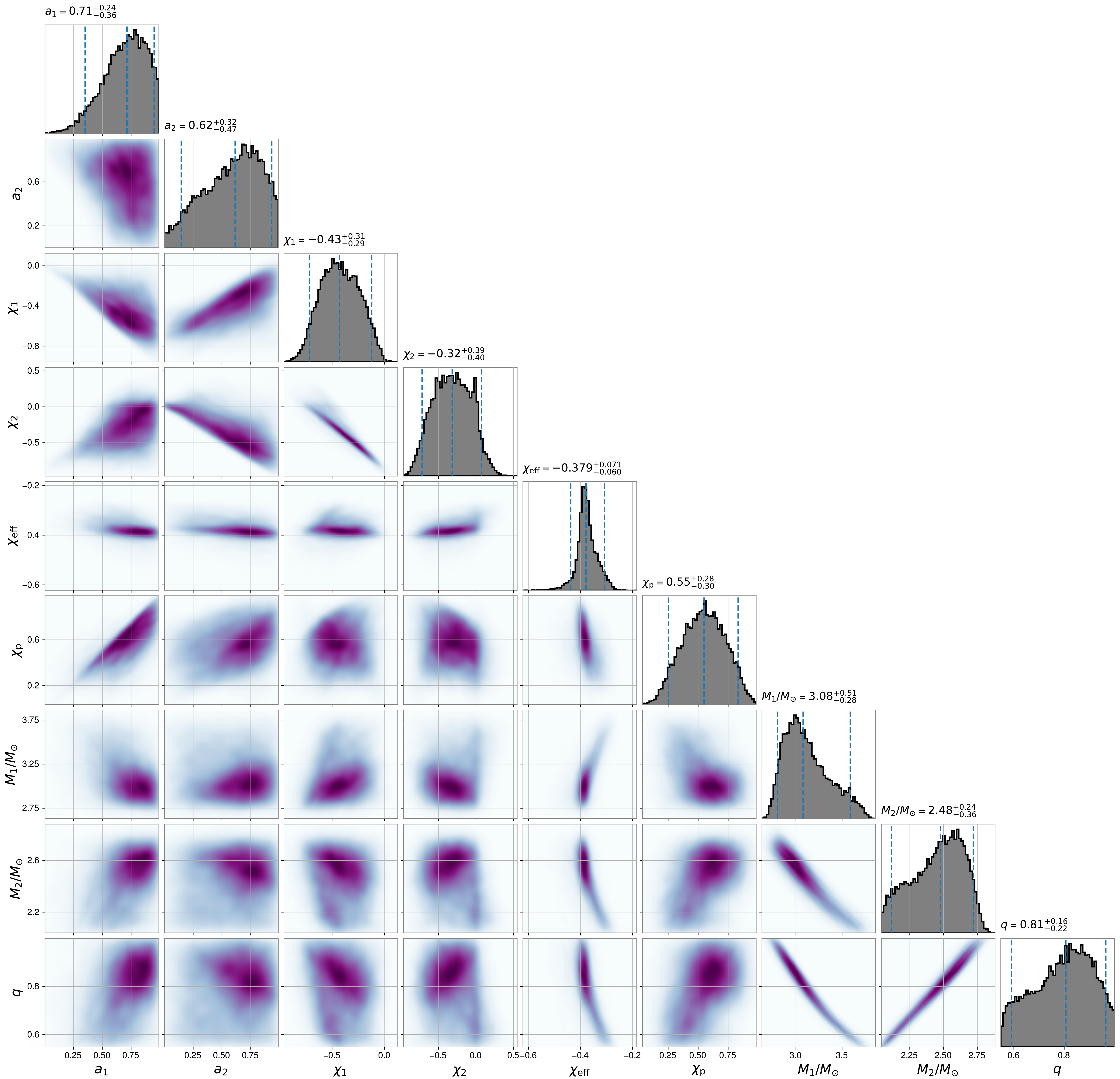}
\caption{Posterior distributions of the physical parameters of GW190426\_152155 on the case of BBH, where we constrain the $M_2 > 2.04M_{\odot}$, and the result is obtained by the \text{IMRPhenomXPHM} waveform. The parameters include the mass ratio ($q$), the effective spin ($\chi_{\rm eff}$), and the effective precession spin ($\chi_{\rm p}$) of the binary system, the spin magnitudes, aligned spins, and the masses of two component objects ($a_1$, $a_2$, $\chi_1$, $\chi_2$, $M_1$, $M_2$).
}
\label{bbh}
\end{figure*}

Motivated by the above results, we further investigate the BBH scenario for GW190426\_152155. We assume that the secondary BH has a mass greater than the heaviest  stable NS. Conservatively, we constrain the source frame secondary mass $M_2>2.04M_{\odot}$  i.e., it is above the 1$\sigma$ lower limit on the mass of PSR J0740+6620 \citep{2020NatAs...4...72C}, and the \text{IMRPhenomXPHM} waveform is adopted to fit the data. The posteriors of the physical parameters are shown in Fig.\ref{bbh}. The masses of two components are ($3.08^{+0.51}_{-0.28}$, $2.48^{+0.24}_{-0.36}$)$M_{\odot}$ respectively, the effective spin parameter $\chi_{\rm eff}$ = $-0.379^{+0.071}_{-0.060}$ with two components' aligned spins ($-0.43^{+0.31}_{-0.29}$, $-0.32^{+0.39}_{-0.40}$), and spin magnitudes ($0.71^{+0.24}_{-0.36}$, $0.62^{+0.32}_{-0.47}$). Additionally, the effective precession spin parameter $\chi_{\rm p} = {\rm Max}( a_1\sin\theta_1, a_2\sin\theta_2(4q+3)/(3q+4))$ is $0.55^{+0.28}_{-0.30}$, which indicates that the spin angular momentum of two compact stars should be misaligned. 

In view of the plausible negative $\chi_{\rm eff}$ and the misalignment of two spin angular momentum, it is reasonable to speculate that GW190426\_152155 may originate from dynamical capture. However, for such low mass BHs,  the dynamical capture process may be inefficient. Moreover, it is unclear whether there is indeed a large population of low mass (i.e., $\sim 3M_\odot$) back holes formed astrophysically. These two cautions should be bear in mind though the GW data alone can not rule out the BBH merger origin of this event.

\subsection{The case of NSBH merger}\label{NSBH}

\begin{figure*}
\centering
\gridline{\fig{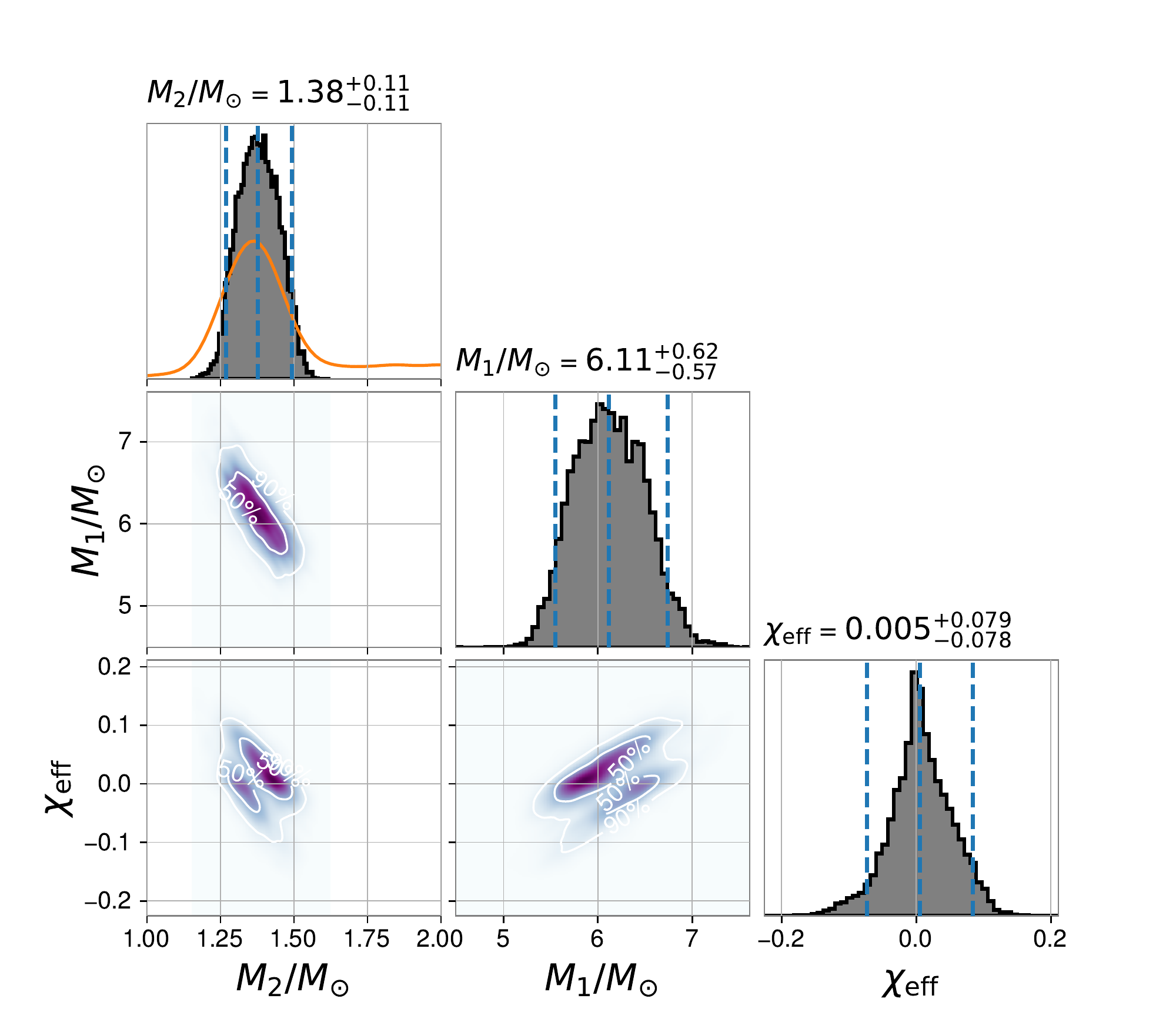}{0.45\textwidth}{(a)}
          \fig{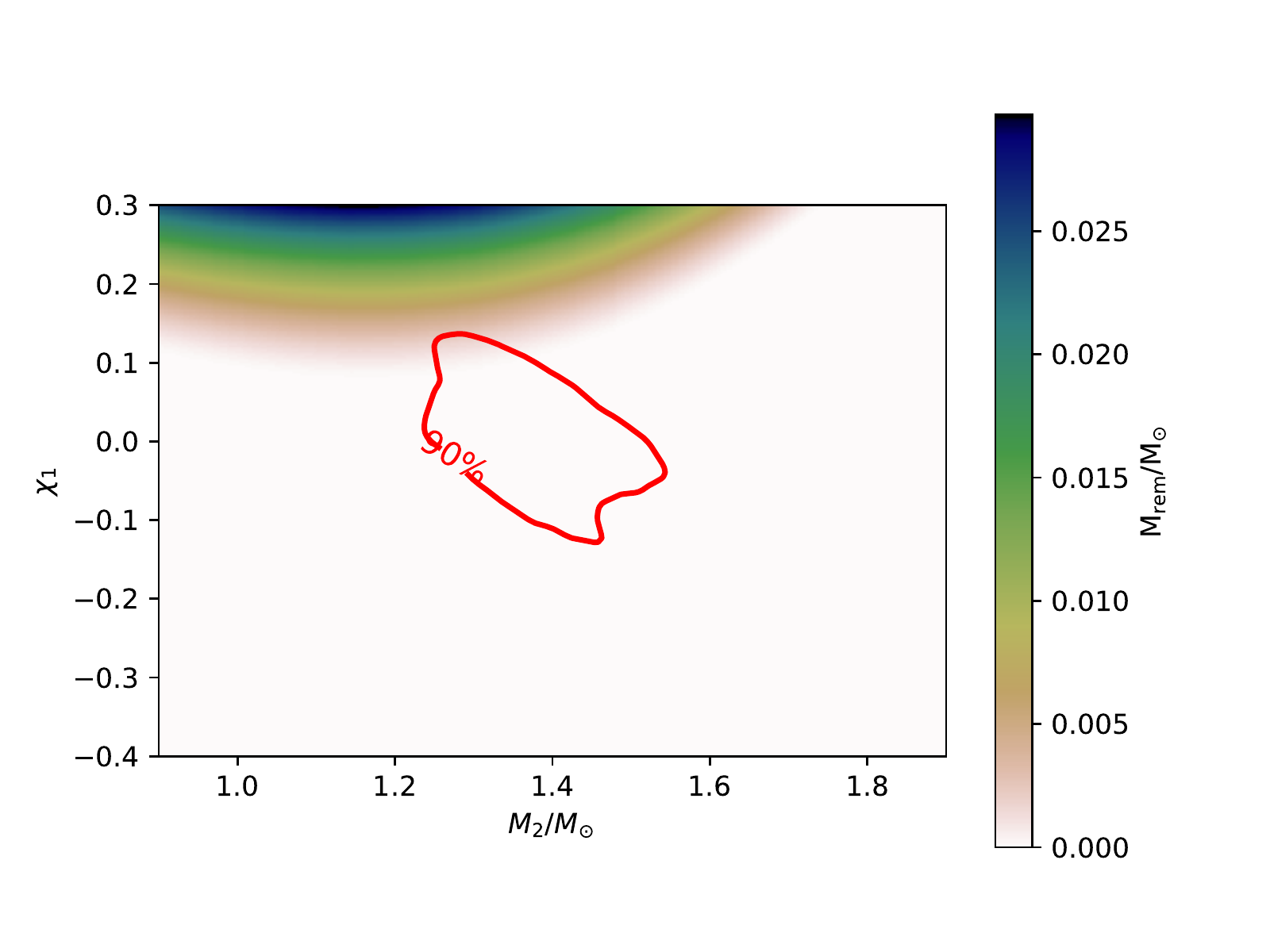}{0.45\textwidth}{(b)}
          }
\caption{Left: the posterior distributions of the source-frame component masses and the effective spin inferred from the population informative prior of NS, the orange line is the prior of the secondary mass. Right: The color bar denotes $M_{\rm rem}$ of the NSBH merger  as a function of the $M_{\rm NS}$ and the $\chi_{\rm BH}$. (We fixed chirp mass of NSBH is as $2.4M_{\odot}$, and using EOS H4 for calculation). The red contours are the and 90\% regions of the posterior distributions obtained from the population informative prior inference.}
\label{m_rem}
\end{figure*}

We further consider a population informative prior for NSBH model, which assumes that $M_2$ belongs to the population of the NSs that with a reasonably measured/constrained mass. We set the prior of the NS mass following \citet{2020PhRvD.102f3006S}, i.e. the source-frame secondary mass has a double Gauss distribution with ($\mu_1$, $\mu_2$, $\sigma_1$, $\sigma_2$) = (1.36, 1.91, 0.09, 0.51)$M_{\odot}$, where $\mu_1$/$\mu_2$, $\sigma_1$/$\sigma_2$ are the mean value and standard deviation of the first/secondary Gaussian component, the fraction of the first Gaussian component is 0.65, and the low and high cut are 0.9 and 2.26 $M_{\odot}$. Then the NSBH model \text{IMRPhenomNSBH} is used to fit the data, and the results are shown in Fig.\ref{m_rem}(a), including the component masses ($M_1$, $M_2$) = ($6.11^{+0.62}_{-0.57}$, $1.38^{+0.11}_{-0.11}$)$M_{\odot}$, and the effective spin $\chi_{\rm eff} = 0.005^{+0.079}_{-0.078}$, we find that the peak of $M_2$ in posterior just matches the first Gaussian component in the prior of $M_2$.
Additionally, since the tidal effect is really weak, we also use \text{IMRPhenomXPHM} to extract the signal from data in NSBH case, while the results are rather similar to that of \text{IMRPhenomNSBH}, i.e. $M_2$ is dominated by its prior.

In the NSBH merger scenario, it is interesting to investigate whether GW190426\_152155 could also generate detectable EM emission. The most relevant information is the mass remaining outside the BH after NSBH merger ($M_{\rm rem}$), which mainly includes the dynamical ejecta ($M_{\rm dyn}$) and the disk mass ($M_{\rm disk}$) of the merger. Following \citet{2016ApJ...825...52K} and \citet{2018PhRvD..98h1501F}, we can use the masses of two component objects ($M_{\rm BH}$, $M_{\rm NS}$), aligned spin of the BH ($\chi_{\rm BH}$), and the tidal deformability of the NS ($\Lambda_{\rm NS}$) to estimate the $M_{\rm rem} = M_{\rm dyn} + M_{\rm disk}$. To illustrate the $M_{\rm rem}$ produced by different $M_{\rm NS}$ and $\chi_{\rm BH}$, we take a typical value of the well constrained source frame chirp mass $\mathcal{M}$ as $2.4M_{\odot}$, then the $M_{\rm BH}$ is a function of $M_{\rm NS}$. Since the tidal deformability of the NS is not constrained, we calculate $\Lambda_{\rm NS}$ as a function of $M_{\rm NS}$  with an EOS. For an optimistic estimation of $M_{\rm rem}$, we adopt a stiff EOS H4 \citep{2006PhRvD..73b4021L,2020CQGra..37d5006A}.
As shown in Fig.\ref{m_rem}(b), the color bar denotes $M_{\rm rem}$, and the red contours represent the and 90\% regions of the posterior of GW190426\_152155 obtained by the population informative prior as mentioned above, we can find the chance of observing a NSBH merger with EM counterparts is rather unpromising for GW190426\_152155, due to the low spin of the BH and hence the very small $M_{\rm rem}$.

\section{Single-detection based merger rate estimation}\label{merger rate}
Whether low mass BBH or NSBH merger does GW190426\_152155 originate from, this kind of events can consider as a new class of compact binary mergers. We estimate the rate of mergers similar to this source, assuming a constant rate per co-moving volume-time element. Following \citet{2018CQGra..35n5009T}, we calculate our surveyed spacetime volume $\left \langle VT \right \rangle$ semi-analytically, where we inject simulated signals to strain data generated by the PSDs of the O1, O2, and O3a and recover them. Following \citet{2019ApJ...882L..24A}, for the O1 and O2, the PSD in each LIGO interferometer is approximated by the Early High Sensitivity curve in \citet{2018LRR....21....3A} (the file can be found in \url{https://dcc.ligo.org/LIGO-P1200087-v46/public}), and for the PSD of O3a is fetched from (\url{https://dcc.ligo.org/LIGO-T2000012/public}). We consider a simulated signal being detected, whose network SNR is greater than 8, this agrees to the lowest SNR among the candidates in O3a. Note that our operation is different from that in \citet{2020ApJ...896L..44A} and \citet{2020ApJ...900L..13A}, where they add simulated signals to data from the O1, O2 and O3a observing runs and assume a once-per-century FAR threshold.

The injection signals are uniformly distributed over co-moving volume and time with randomized orientations, and since the population model for such sources (either NSBH or low mass BBH) was not constructed yet, the source parameters of the signals are prepared by directly drawing from a posterior distribution inferred from the waveform \text{IMRPhenomXPHM} \citep{2003ApJ...584..985K}, this method was also performed in \citet{2020ApJ...896L..44A} and \citet{2020ApJ...900L..13A}. As was introduced in \cite{2020arXiv200602211E} we consider 116/269/183 days of observation with 41\%/46\%/60\% coincident operation of both LIGO's detectors for O1,O2, and O3a. We do not include the Virgo detector for simulation, and this makes few influence on the result. 

From the injection simulation, we find the combined searched spacetime volume $\left \langle VT \right \rangle$ over O1, O2, and O3a is 0.02 $\rm {Gpc}^{3}~{\rm yr}$. We take a Poisson likelihood over the astrophysical rate R and use a Jeffreys prior as p(R) $\propto \rm R^{-0.5}$ to obtain rate posteriors. With the fact that one event is detected, we obtain the merger rate of GW190426\_152155 as $59^{+137}_{-51}~{\rm Gpc}^{-3}~{\rm yr}^{-1}$, and we conclude that the uncertainty in our estimate of the rate density for the class of mergers represented by GW190426\_152155 is primarily dominated by Poisson statistics. We also perform our estimation on the GW190521 and GW190814 for cross check, then obtain $0.08^{+0.17}_{-0.06}~{\rm Gpc}^{-3}~{\rm yr}$ and $13^{+31}_{-11}~{\rm Gpc}^{-3}~{\rm yr} $, which agree with the results in \citet{2020PhRvL.125j1102A} and \citet{2020ApJ...896L..44A} respectively. 

\section{Discussion}\label{diss}
As one of the interesting candidates in O3a Catalog \citep{2020arXiv201014527A}, GW190426\_152155 may be the first GW signal from a NSBH merger. We have performed the Bayesian Inference for GW190426\_152155 using two different priors of mass ratio, and find that the priors can strongly influence the results of the posteriors and we can not rule out a BBH origin. If interpreted as a low mass BBH merger, this binary may have a significantly negative effective spin and an un-ignorable effective precession spin, which may indicate a dynamical capture process.  The cautions are however whether there are indeed a group of such low mass BHs formed astrophysically and the low efficiency of the dynamical capture for the light objects.
In the case of a NSBH origin, we find that this system is hard to produce a bright EM emission because of the rather low spin of the BH. Supposing the signal has an astrophysical origin, we find the merger rate of GW190426\_152155 like systems is $59^{+137}_{-51}~{\rm Gpc}^{-3}~{\rm yr}^{-1}$, which is consistent with the NSBH merger rate suggested in the literature \citep{2010CQGra..27q3001A,2017ApJ...844L..22L}.

Leaving out the high FAR, GW190426\_152155 is indeed an encouraging candidate, while the low SNR of GW190426\_152155 makes it rather hard to study the physical properties of such kind of events. Fortunately, in the O5 run of LIGO/Virgo/KAGARA, the sensitivities of the detectors will be substantially enhanced, for example, the orientation-averaged horizon is about 3 times farther than that in the O3 run (\cite{2018LRR....21....3A}; \url{https://dcc.ligo.org/public/0161/P1900218/002/SummaryForObservers.pdf}), hence more and more such events may be detected. As for the NSBH binary, at that time, the NSBH merger detection rate will be enhanced by a factor of a few tens. With a reasonably large sample, the BH mass function in the merging NSBH systems can be reliably reconstructed \citep{2020ApJ...892...56T,2018ApJ...856..110Y}, which will shed light on the formation or evolutionary paths of these BHs, and the difference between the BH mass functions for the merging NSBH and BBH systems, if significant, would be revealed too.

\acknowledgments

We thank Jin-Liang Jiang for constructive suggestions. This work was supported in part by NSFC (i.e., Funds for Distinguished Young Scholars) under grants of No. 11525313, No. 11921003 and No. 11703098, the Chinese Academy of Sciences via the Strategic Priority Research Program (Grant No. XDB23040000), Key Research Program of Frontier Sciences (No. QYZDJ-SSW-SYS024), and Guangdong Major Project of Basic and Applied Basic Research (Grant No. 2019B030302001). This research has made use of data and software obtained from the Gravitational Wave Open Science Center (\url{https://www.gw-openscience.org}), a service of LIGO Laboratory, the LIGO Scientific Collaboration and the Virgo Collaboration. LIGO is funded by the U.S. National Science Foundation. Virgo is funded by the French Centre National de Recherche Scientifique (CNRS), the Italian Istituto Nazionale della Fisica Nucleare (INFN) and the Dutch Nikhef, with contributions by Polish and Hungarian institutes.\\

\vspace{5mm}


\software{Bilby \citep[version 0.5.5, ascl:1901.011, \url{https://git.ligo.org/lscsoft/bilby/}]{2019ascl.soft01011A},
          PyMultiNest \citep[version 2.6, ascl:1606.005, \url{https://github.com/JohannesBuchner/PyMultiNest}]{2016ascl.soft06005B},
          PyCBC \citep[gwastro/pycbc: PyCBC Release v1.14.1, \url{https://github.com/gwastro/pycbc/tree/v1.14.1}]{2019PASP..131b4503B,2019zndo...3546372N}
          }

\bibliographystyle{aasjournal}
\bibliographystyle{apsrevlong}

\end{document}